\documentclass[preprints,article,accept,moreauthors,pdftex,10pt,a4paper]{mdpi}

\firstpage{1} 
\pubvolume{xx}
\issuenum{1}
\articlenumber{5}
\pubyear{2019}
\copyrightyear{2019}
\history{Received: date; Accepted: date; Published: date}

\usepackage{amsmath,amssymb} 

\Title{The wind dynamics of super-Eddington sources in FRADO}

\Author{Mohammad-Hassan Naddaf $^{1,2}$*\orcidA{}, Bo\.{z}ena Czerny $^{1}$\orcidB{}, Michal Zaja\v{c}ek $^{3,1}$\orcidC{}}

\AuthorNames{Mohammad-Hassan Naddaf, Bo\.{z}ena Czerny, Michal Zaja\v{c}ek}

\address{%
$^{1}$ \quad Center for Theoretical Physics (PAN), Al. Lotnik\'ow 32/46, 02-668 Warsaw, Poland\\
$^{2}$ \quad Nicolaus Copernicus Astronomical Center (PAN), Ul. Bartycka 18, 00-716 Warsaw, Poland\\
$^{3}$ \quad Department of Theoretical Physics and Astrophysics, Faculty of Science, Masaryk University, Kotl\'a\v{r}sk\'a 2, CZ-611\,37 Brno, Czech Republic
}

\corres{Correspondence: naddaf@cft.edu.pl}

\firstnote{All authors contributed equally to this work.} 

\abstract{We perform non-hydrodynamical 2.5D simulations to study the dynamics of material above accretion disk based on the disk radiation pressure acting on dust. We assume a super-accreting underlying disk with the accretion rate of 10 times the Eddington rate with central black hole mass ranging from $10^7$ up to $10^9 M_{\odot}$. Such high accretion rates are characteristic for extreme sources.
We show that for high accretors radiatively dust-driving mechanism based on FRADO model always leads to a massive outflow from the disk surface, and  the failed wind develops only at larger radii. The outflow rate strongly depends on the black hole mass, and in optically-thick energy-driven solution can exceed the accretion rate for masses larger than $10^ 8 M_{\odot}$ but momentum-driven outflow does not exceed the accretion rate even for super-Eddington accretion, therefore not violating the adopted stationarity of the disk. However, even in this case the outflow from the disk implies a strong mechanical feedback.}


\begin{document}


\section{Introduction}

The source of radiation power in Active Galactic Nuclei (AGN) is supplied by the process of accretion of circumnuclear material onto a central supermassive black hole. The black hole mass can range from $10^6$ up to $10^{10} M_{\odot}$. The Eddington luminosity, expressed as $L_{\rm Edd}= 4 \pi G M_{\rm BH} m_{\rm p} c / \sigma_{\rm T} $, sets the limit of the maximum luminosity a spherically symmetric source of radiation can achieve when the force of radiation acting outward and the gravitational force acting inward are balanced. The observed bolometric luminosity for most of AGN is smaller
than the Eddington value \citep{Kollmeier2006}. However, some sources with super-Eddington accretion are reported, they  usually belong to narrow-line seyfert 1 (NLS1) galaxies \citep[e.g.][]{Komossa2006, Jin2016}, type A quasars \citep{sulentic2000}, or weak emission-line quasars \citep{Luo2015}. Other evidence such as the existence of supermasssive black hole of $10^{9} M_{\odot}$ at the redshift of around 7 \citep{Begelman2017}, and tidal disruption events \citep{Rees1988} also hint at the presence of super-Eddington accretion in many situations. Super-Eddington sources were systematically monitored within the SEAMBH project \citep[e.g.][]{DuPu2014,DuPu2015,DuPu2018}. A fraction of highly accreting objects selected from SDSS quasar sample shows super-Eddington accretion, frequently accompanied by blue H$\beta$ line component \citep{negrete2018}.
A study on the accretion rate in four samples of AGN \citep{Collin2004} claims that NLS1s always accrete at Eddington or super-Eddington rates with accretion rates reaching up to $60 \dot{M}_{\rm Edd}$ for Schwarzschild black holes and $300 \dot{M}_{\rm Edd}$ for Kerr ones. They deduce that all galaxies spend at least $0.2\%$ of their life in
the NLS1 phase. A very recent radiative hydrodynamical simulation \citep{Inayoshi2022} found the super-Eddington phase as early stages when the bulge forms and the black hole seed is fed by a massive gaseous disk. The super-Eddington accretion phase was also recently detected in 12 objects within a sample of 42 infrared-luminous galaxy mergers where the super-Eddington accretion is claimed to be an important channel for the black hole growth in infrared-luminous galaxies at all redshifts \citep{Farrah2022}. In general, super-Eddington sources populate the extreme right tail of the Quasar Main Sequence in the optical plane showing the highest ratio of the Fe II pseudo-continuum to H$\beta$ equivalent width and the narrowest emission lines \citep{sulentic2000,marziani2018}, and the 26 sources selected as extreme in this plane show the mean Eddington ratio of 21 \citep{sniegowska2018}. These sources are potentially important for cosmology, since they may serve as standard candles \citep{wang2013b, WangCandles2014, dultzin2020}.

In the super-Eddington regime, most of the accreting matter in excess of the Eddington rate flows out due to radiation pressure \citep{Fukue2004}.
Super-Eddington accretion can produce significant amount of radiation and outflow from the disks which implies a strong feedback from AGN which is crucial in regulating the evolution and structures of their host galaxies \citep{Ciotti2010, Kormendy2013}. Therefore, it is important to investigate the flow dynamics in AGN when the accretion rate is above the Eddington limit. There are many recent numerical simulations aimed at studying the properties of super-Eddington accretion disks \citep[see e.g.,][]{Ohsuga2005, Jiang2014, sadowski2014}. 

In this short article, we aim at studying the dynamics of the dust driven disk winds in super-Eddington regime. Such winds are likely responsible for the formation of the Broad Line Region, and its properties can be strongly affected by the super-Eddington accretion in the disk. We present our simulation results from a 2.5D numerical non-hydrodynamical version \citep{naddaf2021} of the Failed Radiatively Accelerated Dusty Outflow model, so called FRADO \citep{czerny2011}; which is based on radiation pressure from the disk acting on dust developed at the surface layers of the cold disk \citep{czerny2011, czerny2017}. The model was basically developed to address the physics and dynamics of broad line region (hereafter BLR) in AGN. The distinct broad emission lines are generally detected in AGN \cite{2000ApJ...533..631K,2004ApJ...613..682P,2009ApJ...705..199B,2013ApJ...764...47G}, which constitute about $\sim 10\%$ of galactic nuclei. They are characterized by the high accretion rate of the order of $1\,M_{\odot}\,{\rm yr^{-1}}$ when compared with the prevailing quiescent sources, such as the Galactic centre \citep{2017FoPh...47..553E,2021bhns.confE...1K} and M87, which accrete several orders of magnitude below the Eddington limit. The strength of BLR lines points towards the large covering factor, i.e. about $30\%$ of the incident AGN disc photoionizing radiation is intercepted by the BLR clouds. In comparison, the BLR absorption lines are rather rare, which implies the flattened geometry that is likely directly linked to the disc structure. However, the BLR-emitting gas does not lie entirely in the disc plane, since this would result in the too low covering factor, see e.g. BLR reviews \cite{1999agnc.book.....K,2013peag.book.....N}; this is the main motivation behind the FRADO mechanism which assumes the origin of the BLR material within the accretion disk itself, not in the circumnuclear material or stars \citep[see e.g.][]{1997MNRAS.284..967A,1997MNRAS.285..891A}. The model was successful in predicting many observational features of AGN, such as the size of broad line region etc., \citep[see][for more details]{czerny2015, czerny2016, czerny2017}.

The BLR region is very compact, which is the main reason why it took so long to resolve it spatially. Using the Very Large Telescope Interferometer (VLTI) with the instrument GRAVITY, the broad emission line Pa$\alpha$ was kinematically resolved out in the radio quasar 3C273  \cite{2018Natur.563..657G}. These interferometric infrared data clearly showed that the material associated with the broad Pa$\alpha$ line is orbiting the SMBH on sub-parsec scales. The spatial offset of $0.03\,{\rm pc}$ was inferred between the red-shifted and blue-shifted photocenters of Pa$\alpha$ emission line in the direction perpendicular to the previously detected jet. The associated velocity gradient is consistent with the Keplerian rotation of the emitting BLR gas. The interferometric data are well-fitted using the model of a thick disc of BLR cloudlets, which is in a Keplerian rotation around the SMBH of $M_{\rm 3C273}\approx 3\times 10^8\,M_{\odot}$. The determined BLR radius is 150 light days $\sim 0.126\,{\rm pc}$, which is in agreement with the previous reverberation-mapping studies. Two more sources -- NGC 3783 and IRAS 09149-6206 -- observed with the same instrument \citep{GRAVITYIRAS_2020,GRAVITY3783_2021} showed the same pattern using the broad Br$\gamma$ line.

The BLR has been proposed to consist of low- and high-ionization line-emitting regions \citep{1988MNRAS.232..539C}. The low-ionization emission lines (LIL, with an ionization potential $\lesssim$20 eV), where most of the Balmer lines such as $H\beta$, CII $\lambda1336$, and MgII$\lambda2800$ and the pseudo-continuum of FeII belong, are supposed to originate in the mildly ionized, higher density regions $(n>10^{11}\,{\rm cm^{-3}})$. The LIL-emitting material is then likely located close to or directly in the extended parts of the accretion disc with no distinct inflow/outflow signatures. When the accretion rate increases and approaches the Eddington limit, LILs do, however, exhibit the presence of outflows (e.g. \citep{2018A&A...620A.118N}). In contrast, the high-ionization lines (HIL), with an ionization potential $\gtrsim$40 eV, for instance Ly$\alpha$, CIV $\lambda1549$, HeII $\lambda1640$, and NV $\lambda1240$, are associated with the highly ionized, lower density regions $(n < 10^{10}\,{\rm cm^{-3}})$. They form an outflow beyond the disc, which is revealed via the blueward asymmetries and line centroid shifts. However, a part of the HIL-emitting material is positioned geometrically close to the central black hole, as is indicated by the reverberation mapping studies, which indicate the existence of the analogous radius-luminosity relation as for the LILs, at least for the CIV broad line \citep{2021ApJ...915..129K}. Especially already mentioned blueward asymmetries and blushifted peaks with respect to the rest-frame of the HIL reveal the existence of disk winds in AGN \cite{1995ApJ...451..498M, 2009NewAR..53..140G, 2017A&A...608A.122S}. The properties of LILs show that gravitational and radiation forces both affect the dynamics of the BLR \citep{2018NatAs...2...63M}. The presence of disk winds is also consistent with broad absorption features in the UV and X-ray domains of AGN spectra, which are generally blueshifted \cite{2004ApJ...616..688P, 2010MNRAS.408.1396S}. We were also able to catch the outflow feature of the BLR with the 2.5D FRADO model \citep{naddaf2021, Naddaf2022}, and the blushifted peak of the line profile as well \citep{Naddaf2022}. However, in essence our model applies to the LIL part of the BLR, and the radial distance from the black hole supports that \citep{naddaf2020}, while the HIL line delays are considerably shorter.

We structure the paper as follows. In Section 2 we introduce the FRADO model and present the results of simulations of the dynamics by 2.5D FRADO for the super-Eddington case. We present the analysis of the outflow rate from the disk in the our simulations in Section 3, followed by discussion in Section 4 with the observational implications of our results included. A summary of the study is then presented in Section 5.

\section{Dynamics in 2.5D FRADO model}

The basic analytical FRADO model developed a decade ago \citep{czerny2011, czerny2015, czerny2016, czerny2017} addresses the dynamics of dusty material lifted from the disk surface due to the radiation pressure of the accretion disk acting on dust. The motion of material is in the form of a failed wind, described in one dimension, i.e. vertical direction, without the orbital motion included. 
In this model, the material composed of dust and gas in the form of clumps developed at the surface layers of the cold accretion disk \citep{rees1969, dong2008}, is launched by the local radiation flux of the accretion disk itself. Once the clumps reach high altitudes, they are exposed to the strong irradiation from the central disk so that they lose their dust content. The remaining gaseous clumps then follow a ballistic motion in the gravitational potential of the central black hole and finally fall back to the disk surface.

We then developed the 2.5D (azimuthally symmetric) numerical version of the model. It is a nonhydrodynamical single-cloud approach to the dynamics of clumps. Compared to the 1D model, it is enhanced with the realistic
description of the dust opacities \citep{rollig2013} interacting with the radiation field; and the intense central radiation is geometrically shielded for the early lifted clumps, known as shielding effect \citep{murray1995, risaliti2010}. We previously introduced our 2.5D model \citep[see][for more details]{naddaf2021}, and tested it against the measured size of BLR \citep{naddaf2020}, and with calculation of emission line profiles \citep{Naddaf2022}. The collisions with the disk surface based on 2.5D FRADO model are also of interest since they can also give rise to non-thermal X-ray and gamma-ray emission due to the particle acceleration to relativistic velocities in strong shocks \citep{2022ApJ...931...39M}. Those works considered the values of the luminosity up to the Eddington value.

\begin{figure*}
	\centering
	\includegraphics[scale=0.37]{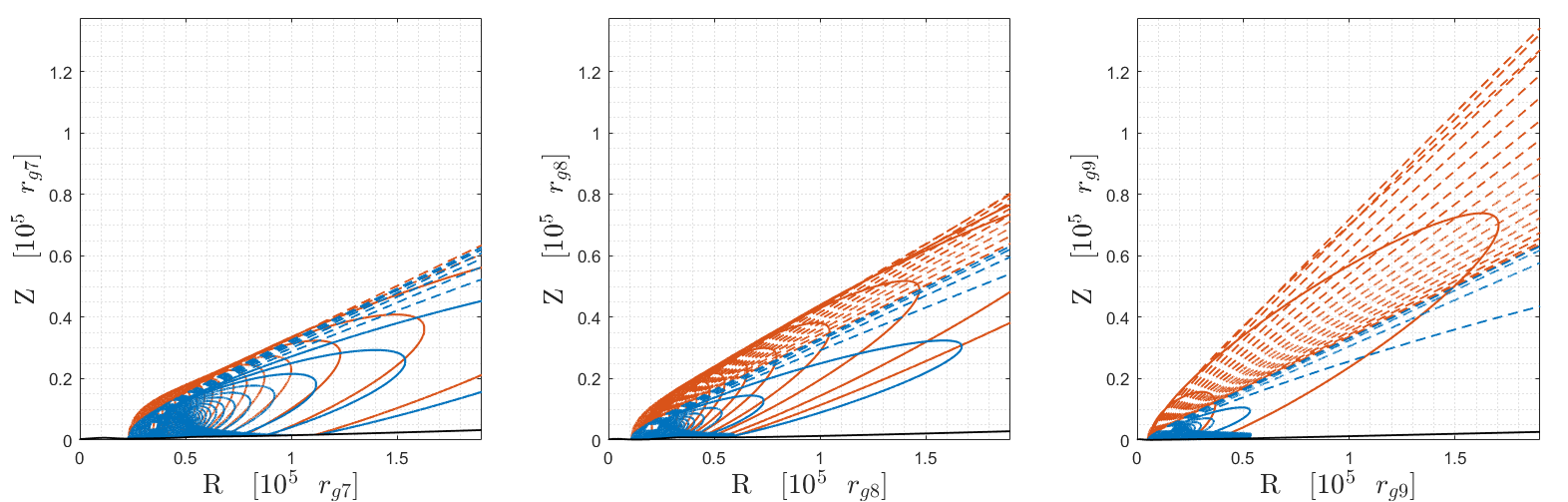}
	\caption{The result of the simulation of the trajectories of clouds lifted from the disk surface with the central black hole mass of $10^7 M_{\odot}$ (left), $10^8 M_{\odot}$ (middle), and $10^9 M_{\odot}$ (right). The accretion rate of the source is fixed at 10 $\dot{M}_{\rm Edd}$. The blue and red colors mark the dusty and dustless clouds, respectively. Solid and dashed lines represent the trajectory of failed and escaping clouds, respectively. The black solid line represents the disk surface.}
	\label{fig:trajects}
\end{figure*}

Here in this work, therefore, we decided to perform the simulations for a super-Eddington accretion disk with the accretion rate of 10 times the Eddington limit. For the configuration of the model, one can refer to our main paper \citep{naddaf2021}. The clumpy material initially at circular Keplerian motion around the central black hole and an initial zero vertical velocity is lifted up due to radiation pressure acting on dust content of the clump. The equation of motion of the clump can be qualitatively described as:

\begin{equation}
\textbf{a}^{\rm tot} =
- \textbf{a}^{\rm grav}
+ \textbf{a}^{\rm rad}\ (\textbf{r}, \Psi, M_{\mathrm{BH}}, \dot{M},
\kappa^{\mathrm{ext}}_{\lambda}, \textbf{S}, T_{\mathrm{subl.}})\
\label{eq:motion}
\end{equation}
where $M_{\mathrm{BH}}$ stands for the black hole mass, $\psi$ is the dust-to-gas mass ratio, $\kappa^{\mathrm{ext}}_{\lambda}$ is the dust opacity as a function of wavelength, $\dot{M}$ stands for the accretion rate, $T_{\mathrm{sub}}$ is the dust sublimation temperature, and $\textbf{S}$ the shielding-implied effective area of the disk contributing to the radiative force on the clump, and $\textbf{r}$ represents the position vector of the clump. The second term which is the radiative component of acceleration is in play as long as the dust temperature is below that of sublimation.

We then solve the equation of motion for a range of initial parameters with the fixed accretion rate of 10 times the Eddington value, based on which we aim to study and analyse how the model works in the super-Eddington regime. For this project, the modification of the disk structure due to advection \citep{abramowicz1988} are unimportant since we mostly concentrate on the disk structure at larger radii.

For this purpose we considered 3 values of the central black hole mass of $10^7$, $10^8$, and $10^9 M_{\odot}$. The initial dust content of the clumps is considered to be the same as the Solar value, the accretion efficiency is set at 0.1, and the dust sublimation temperature is fixed at $1500$ Kelvin.

Figure~\ref{fig:trajects} shows the results of simulations ,in which only a fraction of trajectories out of a very dense simulated set of initial radii are displayed for the matter of a better visibility. Each line displays the path motion of a clump launched from a certain radius. The clumps at the time of launching are in local circular Keplerian motion within the disk around the central black hole, with no initial vertical velocity. A clump is then launched due to the local disk radiation pressure, and it starts to move under the radiative dust-driving force and gravitational force of the central black hole. Depending on the launching radius, the clump may follow a closed trajectory path and come back to the surface of the disk which is shown by solid lines, or it can escape to infinity shown by dashed lines. The trajectories color-coded in blue show the path of motion of dusty clumps, while those marked in red are reserved for dustless ones. The launched clouds are initially dusty. Once reaching a certain height above the disk (so called {\it sublimation location}), they are exposed to the intense central disk irradiation, which results in the loss of their dust content. Those losing their dust follow a ballistic motion under the gravitational influence of the central black hole.

As can be seen from Figure \ref{fig:trajects}, the clumps lifted from the disk within a range of radii (we call it {\it escaping zone}) can reach high enough velocities to escape the gravitational potential to infinity in the form of a stream of outflowing material. Considering the models calculated for the Eddington rate or lower, this escaping stream was never present for black hole masses characteristic for NLS1, of the order of $10^7 M_{\odot}$ and for a solar metallicity \citep{Naddaf2022}. According to our model, sources with a strong outflowing stream of material should exhibit a blue asymmetry in the line profile at viewing angles above 45 degrees. We can compare this prediction with the exceptional source I Zw 1, for which the black hole mass of $9.30_{-1.38}^{+1.26} \times 10^6 M_{\odot} $ was determined using the reverberation mapping \citep{Huang2019}. This object does not show such a blue asymmetry in H$\beta$ line, although it is highly super-Eddington \citep{Huang2019}. Therefore, we conclude that the viewing angle cannot be larger than $\sim 30$ degrees, i.e. it should be smaller than $\sim 40$ deg adopted by \citet{Huang2019} in their mass derivation. Unfortunately, there are no observational constraints for the viewing angle of this source.

The opening angle of the outflow, $\Delta \theta$, inferred from our simulation results, is $8.1, 10.9, 23.9$ degrees for the 3 adopted black hole masses from the smallest to the largest value, respectively. It clearly indicates that with the increase in the black hole mass, the outflow covers a larger patch of the sky above the accretion disk.

We found that with the increase in the black hole mass, the onset of the escaping zone, $R_{\rm zone}^{\rm in}$, shifts inwards as $32\,507~r_{g7}$, $16\,406~r_{g8}$, and $5\,836~ r_{g9}$ (using the units of the corresponding black hole gravitational radius). Its width, $\Delta R_{\rm zone}$, gets drastically broader as $3\,045~ r_{g7}$, $5\,843~ r_{g8}$, and $9\,193~ r_{g9}$. The width of the escaping zone may imply that the outflow gets more massive as the black hole mass increases. We will confirm this trend by estimating the amount of mass lost due to the outflow from the accretion disk, which is the subject of the following section.

It should be noted that, as we previously addressed \citep{naddaf2021}, the interactions among clouds in those trajectories crossing each other are rare so that the clouds may experience at most one collision during a full orbit. Therefore, we can safely neglect the adjustment of a given trajectory due to the adjacent trajectories without the loss of significance of our results.

\section{Mass-loss rate due to dusty-driven outflow}
\label{sec:outflow}

As a result of our simulations, the clouds launched within the {\it escaping zone} go to infinity escaping the radial domain of computation in our simulations and form a fast funnel-shaped stream of material inclined with respect to the disk equatorial plane. The formation of the outflowing stream of material in our simulations is consistent with the empirical picture of AGN based on direct observational arguments \citep{elvis2000, Chelouche2019,hoenig2019}. This feature was previously captured by others in non-hydrodynamical \citep{risaliti2010, nomura2013}, and hydrodynamical \citep{proga2000, proga2004, sim2010, higginbottom2014, nomura2020} simulations. However, the  inner radius of the escaping outflow was typically closer in to the black hole due to the line-driving force. Such an outflow is a good description of the HIL part of the BLR, where lines like CIV and He II are formed, and outflow velocities are faster, but the outflow is less massive than the dusty-driven outflow in the LIL part of the BLR. The 2.5D FRADO is the only method which incorporates appropriately the role of dust into the computations and gives rise to the formation of an outflow in the dust-dominated region of the accretion disk \citep{naddaf2021, Naddaf2022, Naddaf2022b}. 

The outflow can affect the accretion process due to mass loss, and the intensity of the mass loss during the outflow is connected with the strength of the feedback from the source to the host galaxy. It is hence important to estimate the mass-loss rate from the disk based on our simulations for the proposed grid of initial conditions. Our numerical model predicts the dynamics, i.e. the velocity field, but not the density field. We thus consider two extreme assumptions about the launching wind, i.e. optically thick and optically thin approximation for the launched material, which soon later forms the clumps \citep[see][for more details]{czerny2017, Naddaf2022b}. These two assumptions provide us with an upper limit and a lower limit for the mass-loss rate, respectively. In the optically-thin approximation, similar to stellar winds, it is assumed that the momentum of radiation is fully transferred to outflow. In this conservative case, most of the energy of the radiation field is not used, so it sets the lower limit of the mass-loss rate. Such a regime in stars is characteristic of hot stars, e.g. O and B stars from the main sequence. The other case corresponding to the upper limit of the mass-loss rate assumes that the whole radiation energy is transferred to the optically thick clumps through multiple scatterings. Such a regime in stars offers a better approximation for the winds from cool, evolved stars like stars on an asymptotic giant branch (AGB) as well as from Wolf–Rayet stars.

\begin{figure*}
	\centering
	\includegraphics[scale=0.37]{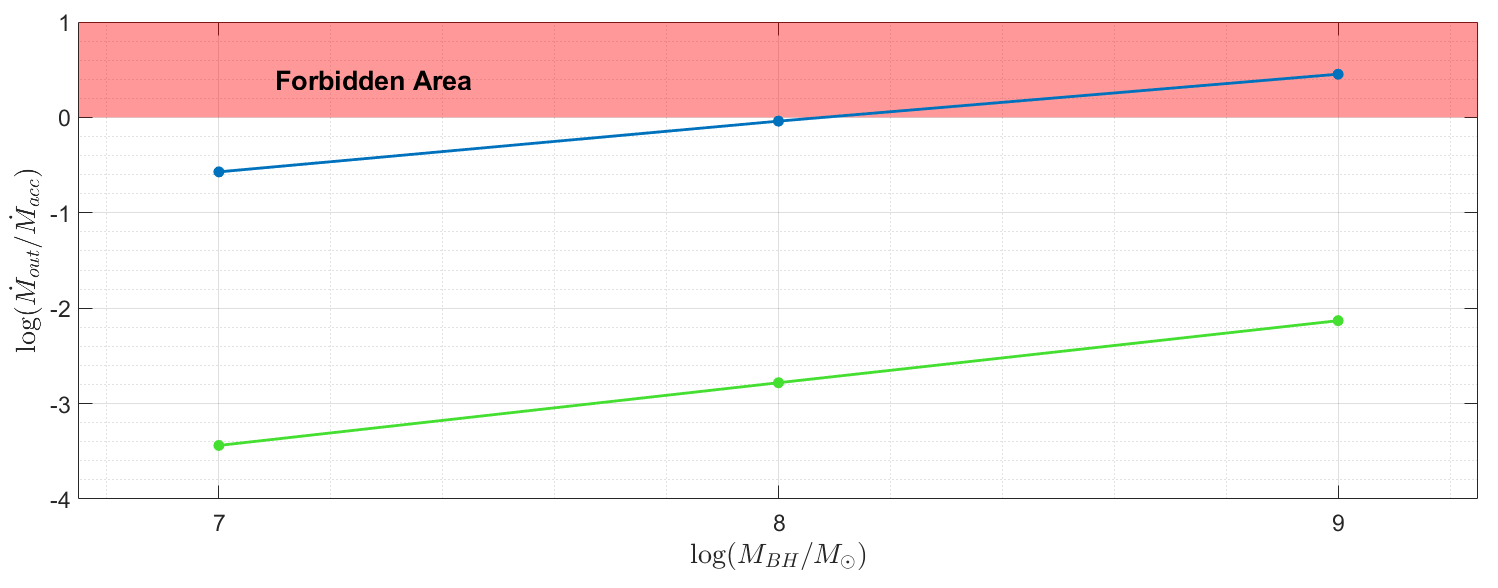}
	\caption{Mass-loss rate divided by the accretion rate of the source for the optically thick (blue line) and the optically thin (green line) approximations. The accretion rate of the source is super-Eddington and is set to 10 $\dot{M}_{\rm Edd}$. The region marked as {\it Forbidden Area} means the outflow rate cannot exceed the accretion rate of the source itself.}
	\label{fig:massloss}
\end{figure*}

Figure~\ref{fig:massloss} shows the results of the radially integrated mass-loss rate inferred from the computations normalized by the accretion rate of the source set initially to 10 times the Eddington limit. The two solid lines in blue and green set the upper and lower limits of the mass-loss rate for optically thick and optically thin approximations, respectively. As it can be seen from the figure, the massive outflow for the optically thick approximation is comparable to what is seen in broad absorption line quasars \citep{borguet2013, chamberlain2015, choi2020}; however, the accretion rate in such sources is not likely that much high. This is the extreme upper limit for the mass-loss rate but the material at large radii of the accretion disk is expected to be not fully but moderately thick. There is a logical restriction for the amount of outflowing material represented in the figure as the {\it Forbidden Area} expressing that the outflow rate cannot be larger than the accretion rate. 

\begin{figure*}
	\centering
	\includegraphics[scale=0.37]{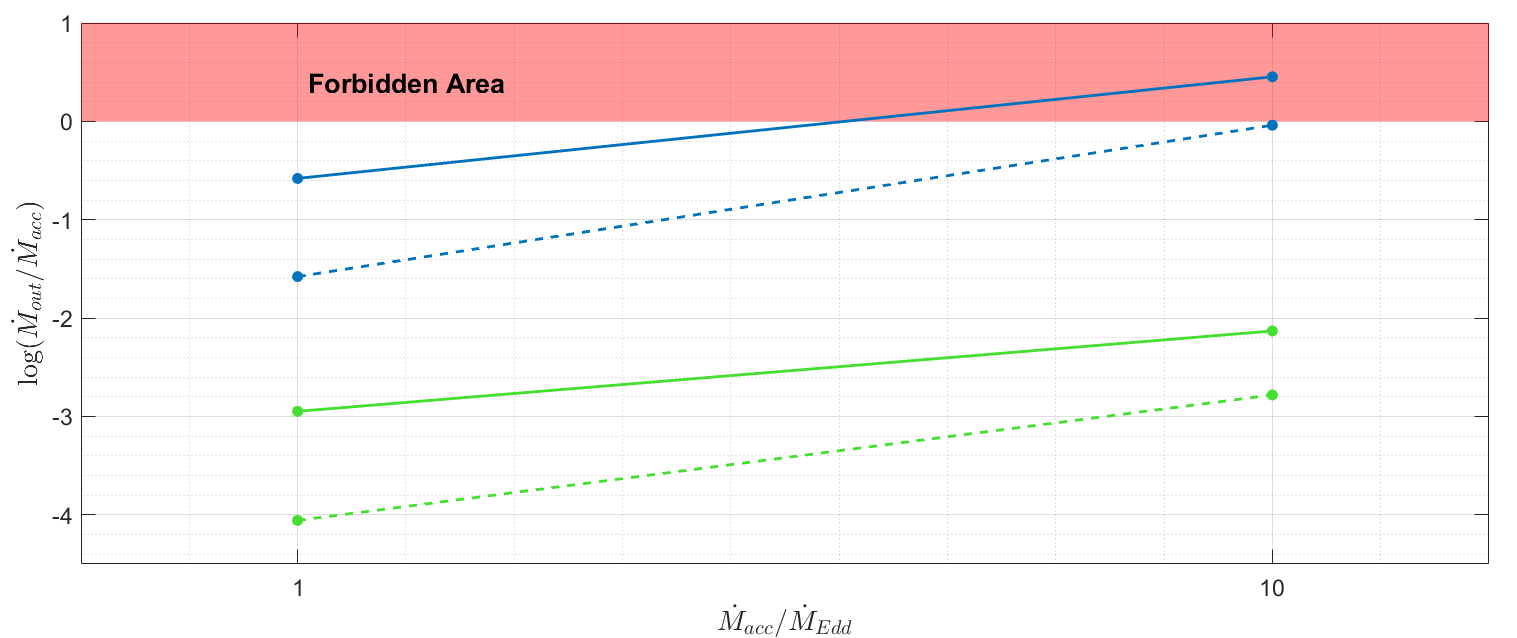}
	\caption{Normalized mass-loss rate for the optically thick (in blue) and the optically thin (in green) approximations; dashed and solid lines correspond to black-hole masses of $10^{8}\,M_{\odot}$ and $10^{9}\,M_{\odot}$, respectively. The accretion rate of the source is expressed in Eddington units. The region marked as {\it Forbidden Area} means the outflow rate cannot exceed the accretion rate of the source itself.}
	\label{fig:massloss_mdot}
\end{figure*}

The Figure~\ref{fig:massloss_mdot} shows the mass loss rate against accretion rate of the source; the full table of results for a range of black hole masses, accretion rates, and metallicities will be available (M. H. Naddaf et al., on BAL quasars, in preparation). The outflow for the super-Eddington accretion rate can be massive but it is important to note that the considerable outflow does not start sharply when the Eddington ratio is crossed, as illustrated in Figure~\ref{fig:massloss_mdot}. This is because the outflow is driven by the radiation pressure acting on dust, so the electron scattering (used in the definition of the Eddington rate) is not relevant here. In addition, the dust to electron scattering opacity is a function of the radial as well as vertical location of the cloud, since the opacity is calculated locally at each cloud position by convolving the locally available radiation flux, including its spectral shape, with the wavelength-dependent cross-section for dust scattering and absorption, as described in detail in \citet{naddaf2020}. The radial distance of the cloud determines the local disk temperature close to the disk surface while the cloud height determines the radial extension of the disk seen by the cloud, and this includes the amount of UV radiation from the central regions. Additionally, the outflow starts at the calculated position of the disk surface which also depends both on the Eddington rate and on the black hole mass. Net outflow, finally, comes from the radial integration of the local outflow rate and is measured in g s$^{-1}$ cm$^{-2}$. This is why the dependence includes, for example, also the black hole mass, and not just the accretion rate or the Eddington rate - both parameters are important. The trend, however, is always monotonic - higher mass and/or higher Eddington rate lead to a stronger outflow. For the Eddington ratio of 0.1 and the black hole mass of $10^8 - 10^9 M_{\odot}$ only failed wind develops, and this is why our dependency shown in Figure~\ref{fig:massloss_mdot} starts at the Eddington ratio of 1.0.

We performed the simulations for the models with the dust content equivalent to solar metallicity, therefore much higher mass-loss rates can be expected if a larger value of metallicity is initially adopted \citep[see e.g.,][]{Naddaf2022b}. The metallicity in AGN is reported in many studies to be much higher than that of the solar value \citep{hamann1992, warner2002, shangguan2018, baskin2018, sniegowska2020}. This would imply (within the framework of our model) that the outflow from the cold disk due to radiation pressure on dust can easily exceed the accretion rate of the disk itself, which would violate the assumption of a stationary disk with a constant accretion rate.

\section{Discussion}\label{sec:discussion}

We performed computations of the outflow from an AGN accretion disk under the action of the radiation-pressure force. We show that for super-Eddington sources considered in this paper the outflow is likely very strong, higher for systems with higher black hole masses. We basically considered the solar metallicity of the wind material, but supersolar metallicity would yield even more vigorous outflow. Such an outflow can exceed the assumed inflow rate, at which point our model, assuming a stationary disk with a constant accretion rate does not provide a self-consistent description of the process anymore.

In principle, we could modify the model to account for the outflow of the BLR material (and the inflow since some clumps return to the accretion disk!) by solving a continuity equation iteratively. 
We would require in most cases much higher accretion rates at the outer edge of the disk than what is finally accreting onto a black hole since most of the material escapes without returning to the disk. However, such iterations are not easy to perform since the rise of the accretion rate in the first iteration shifts the entire escaping zone outwards, completely redistributing the velocity fields. It is not clear that the method will easily converge, it will be certainly very time-consuming, and it is currently beyond the scope of the current article.

On the other hand, it is important to make such an effort in the future. The observed outflows in AGN can be very massive but their launching radii are not well constrained. The best studied of the AGN outflow velocity fields come from CO kinematics \citep[e.g.][]{cicone2014}. However, occasionally, we see there signs of fuelling resolved at 10-20 pc, rather than the outflow \citep[for a review, see][]{combes2022}. Less luminous sources like Seyfert 2 galaxy NGC 1433 are better resolved and mapped with the ALMA down to tens of parsecs \citep{combes2013}, which shows the outflow rate of the order of $7 M_{\odot}$yr$^{-1}$ but possibly connected to starburst and/or jet driving, and the outflow velocities are small ($\sim 200$ km s$^{-1}$) in comparison with expectations from our model ($> 10^3$ km s$^{-1}$). Perhaps more promising in the context of FRADO launching are constraints from the UV and X-ray observations. For example, X-ray studies of winds in MCG-03-58-007 with XMM-Newton and NuSTAR by \citet{braito2022} revealed wind velocities of the order of 0.07 to 0.2 c, which are varying in time, so clearly originating closer to the nucleus, in the accretion disk. The source is, however, heavily absorbed Seyfert 2 galaxy, so no information on the BLR and on the true source bolometric luminosity are then available. However, it may point towards the strong outflow being non-stationary, which would make the future modelling rather complex. Also constraints from the Broad Absorption Line (BAL) quasars may be useful, if these sources are indeed viewed along the outflowing stream (M. H. Naddaf et al., in preparation).

\section{Summary}\label{sec:summary}

We presented the results of non-hydrodynamical simulations of the dynamics of winds in super-Eddington accretion disks based on the FRADO model for a range of masses of the central supermassive black hole, specifically ranging from $10^7\,M_{\odot}$ up to $10^9\,M_{\odot}$. We found that the radiation pressure acting on dust can lead to the formation of funnel-shaped massive winds from the surface of the accretion disk. We calculated the amount of outflowing material in such cases and found that the strength of the outflow and consequently the strength of the mechanical feedback from the source to the host galaxy gets progressively larger with the increase in the central black-hole mass. The simulations were performed for the dust content of an AGN equivalent to the solar metallicity, hence more vigorous outflows can be expected for the cases with supersolar metallicities, which is highly likely in AGN.

\vspace{6pt} 



\authorcontributions{Authors have contributed equally to this work.}


\acknowledgments{The project was partially supported by the Polish Funding Agency National Science Centre, project 2017/26/A/ST9/00756 (MAESTRO 9), and MNiSW grant DIR/WK/2018/12. MZ acknowledges the financial support of the GA\v{C}R EXPRO grant No. 21-13491X ``Exploring the Hot Universe and Understanding Cosmic Feedback". In addition, MZ is grateful for the support by the Czech-Polish mobility program (MSMT 8J20PL037) and by the NAWA grant under agreement PPN/BCZ/2019/1/00069.}

\externalbibliography{yes}
\bibliography{naddaf}



\end{document}